\documentclass[structabstract]{aa}  
\usepackage{graphicx}
\usepackage{txfonts}
\usepackage{natbib}
\bibpunct{(}{)}{;}{a}{}{,} % to follow the A&A style
\newcommand{\be}{\begin{displaymath}}
\newcommand{\ee}{\end{displaymath}}
\begin{document}
  \title{Bayesian timing analysis of giant flare of SGR 1806$-$20 by \emph{RXTE PCA}}

   \author{V. Hambaryan
          \inst{1}
%          \inst{1}\fnmsep\thanks{Valeri Hambaryan, e-mail:~vvh@astro.uni-jena.de}
          \and
	  R. Neuh\"auser
          \inst{1}
          \and
	  K.D. Kokkotas
          \inst{2}
          }

   \institute{Astrophysikalisches Institut und Universit\"ats-Sternwarte, Universit\"at Jena, 07745 Jena, Germany\\
              \email{vvh@astro.uni-jena.de}
         \and
             Theoretical Astrophysics, Eberhard-Karls-Universit\"at T\"ubingen, 72076 T\"ubingen, Germany
             }

   \date{Received ; accepted }

% \abstract{}{}{}{}{} 
% 5 {} token are mandatory
 
  \abstract
{}
   {By detecting high frequency quasi-periodic oscillations (QPOs) and estimating frequencies of them 
during the decaying tail of giant flares from Soft Gamma-ray Repeaters (SGRs) useful constraints for 
the equation of state (EoS) of superdense matter may be obtained via comparison with theoretical 
predictions of eigenfrequencies.}
   {We used the data collected by the Rossi X-Ray Timing Explorer \emph{(RXTE/XTE)} 
Proportional Counter Array \emph{(PCA)} of a giant flare of SGR\, 1806$-$20 on 2004 Dec 27 and applied 
a  Bayesian periodicity detection method (Gregory \& Loredo, 1992) for the search of oscillations of 
transient nature.}
   {In addition to the already detected frequencies, we found a few new frequencies 
($\mathrm{f}_{QPOs} \sim 16.9, 21.4, 36.4, 59.0, 116.3 \mathrm{Hz}$) of oscillations
predicted by Colaiuda et al. (2009) based on the $APR_{14}$ EoS (Akmal et al., 1998) for SGR\, 1806$-$20.}
   {}

%\authorrunning{Hambaryan et al. Bayesian timing giant flare of SGR 1806$-$20}
\titlerunning{Bayesian timing of giant flare of SGR 1806$-$20}
   \maketitle
%
%________________________________________________________________

\section{Introduction}

  The study of periods of activity of Soft Gamma-ray Repeaters 
\citep[SGRs, for a recent review see ][]{2008A&ARv..15..225M} 
showing recurrent bursts with sub-second
duration and much more extreme events known as giant flares (on rare occasions),
may have an important input to our understanding of neutron stars (NSs). 

In particular, the detection and analysis of the presence of quasiperiodic
oscillations (QPOs), up to a few kHz, has triggered a
number of theoretical studies for prediction and direct comparison
with various equations of state for superdense matter. These
oscillations have been interpreted initially as torsional oscillations of
the crust \citep{SA2007,SKS2007}. While later the Alfv\'en 
oscillations of the fluid core have been taken into account 
\citep{2007MNRAS.377..159L,2008MNRAS.385L...5S,CBK2009,DSF2009,2010arXiv1006.0348V,2010arXiv1007.0856G,CK2010Pre}.
These studies suggested how the observations
can constrain both the mass, the radius, the thickness of the crust
and the strength of magnetic field of NSs.
In particular, the timing analysis of the decaying tail of the unprecedent giant flare of 
SGR 1806$-$20 on  2004 Dec 27, allowed to detect QPO frequencies 
approximately at 18, 26, 30, 92, 150, 625, and 
1840 Hz \citep{I2005,WS2006a,SW2006,WS2006b} in different 
time intervals, different rotation phase 
and different amplitudes of oscillations, by means of computation and analysis of the 
averaged power spectrum.

Here, we present the results of a Bayesian approach of timing analysis, 
of the giant flare data set of SGR\, 1806$-$20 registered on 2004 Dec 27 
by \emph{RXTE PCA}, for detecting a periodic signal of unknown 
shape and period developed by \cite{GL1992} \citep[for its sensitivity and advantages see further and also][]{GL1996}.

\section{Observational data and results of analysis}

\begin{figure}
   \centering
   \includegraphics[width=8.6cm]{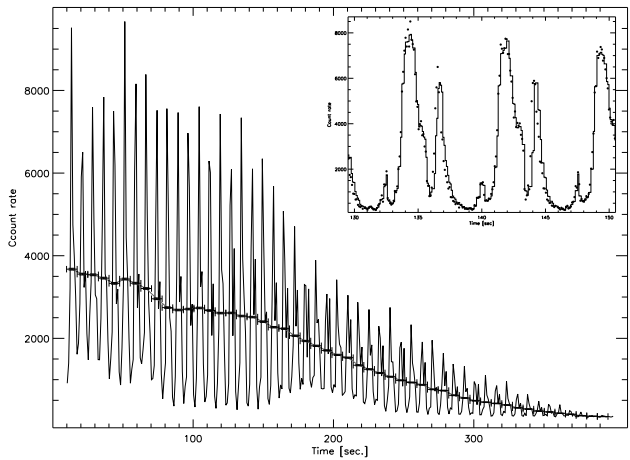}
      \caption{Light curve of the {\it RXTE} giant flare of SGR 1806$-$20 observed on 27th of December 2004. 
The general decay of the giant flare with a bumpy structure on top of it, namely a strongly periodic 
signal due to rotation (7.56 sec). 
The inset panel shows the rotational modulated light curve (filled circles) together 
with fitted piecewise constant model \citep[solid line,][]{Hutter:07pcregx}, 
shown here only for 2.5 rotational cycles, with a very complex light curve structure 
(see text for details).
              }
         \label{lc}
   \end{figure}

The giant flare of SGR\, 1806$-$20
has been observed by many space based missions.
The data recorded by \emph{RXTE PCA} instrument, consisting of 
five Xenon-filled detectors covering the energy range 2$-$50 keV, used
the configuration GoodXenon, which records all good events 
detected in the Xenon chamber with full timing accuracy of 1 $\mu s$. 
Publicly available data were retrieved by the 
\emph{XTE} Data Finder (XDF) user interface \citep{1997ASPC..125..275R}.
Event data files have been created and photon arrival times were 
corrected for the solar barycenter
using scripts provided in the package \emph{XTE ftools}.
The data set consists of 698770 registered photons. However, 
during the initial intensive spike phase of the giant flare the detectors were saturated.
For that reason, in our analysis we use the data after $\sim$~8.9sec of the flare onset, consisting of
$\sim$~650000 photon arrival times, clearly covering 51 rotational cycles 
of SGR\, 1806$-$20 (see Fig.~\ref{lc}).

It is clear that observational detection and parameter estimation of QPO frequencies 
may play a crucial role for testing any theory predicting eigenfrequencies of 
the neutron star. 
In this connection, timing analysis of complex flare data set of SGR\, 1806$-$20, 
with the aim of QPO detection, may be divided into several mutually connected, 
challenging problems.  They include the significance of quasi-periodic signal detection and 
parameter estimation with high precision. Indeed, 
the decaying tail, of the giant flare of SGR\, 1806$-$20, 
itself has a bumpy structure,  very complex 
light curve shape modulated by rotation of the neutron star (see Fig.~\ref{lc}).

The most widely-used procedure for detection of QPOs 
is an analysis of the power spectrum calculated from the 
Fast Fourier transform (FFT) of {\it uniformly} sampled data. Application of the 
FFT to unevenly spaced arrival time series data requires binning of the data to produce {\it equally} 
spaced samples. Binning is a subjective procedure; the choice of the bin width and edges  
can affect the apparent significance of a detection and limits sensitivity on short time scales.
Moreover, the presence of the so-called red-noise (low frequency signal in the data, i.e. 
large calculated power can appear at low frequency because of long-timescale features of the 
data and at high frequencies from higher harmonics of complex shape strong periodic signal),
may cause certain problems with the interpretation of the results \citep[e.g. ][]{brett88}.
In the previous timing analysis of the giant flare data set of SGR 1806$-$20 by \cite{I2005} and \cite{WS2006a}, 
an averaged power spectrum was considered. Namely, for each rotational cycle 
or certain rotational phase interval of SGR 1806$-$20 
an independent classical power spectrum was determined depending on the phase of rotation and 
decaying tail of the giant flare, which were co-added and averaged subsequently. 
This approach divides the data set into small time intervals, which automatically reduces 
the significance of any periodicity detection of transient nature 
and may cause problems related to the windowing of the data sets 
\citep{1989tns..conf...27V,2001SPIE.4477..123G}. 
The complex shape of the light curve (rotational modulation and decaying tail of flare light curves, 
Fig.\ref{lc}) is not taken into account in a proper way, i.e. assuming that during small 
time/phase intervals there are no significant intrinsic variations in the data subsets.
Moreover, in some circumstances FFT may fail to detect the periodic signal 
\citep[see e.g.][]{brett88,GL1996}. As explicitly shown by \cite{Jaynes87} 
\citep[see also,][]{brett88,2001AIPC..568..241B,2005blda.book.....G}
the probability for the frequency of a periodic sinusoidal signal is given approximately by 

\begin{equation}
p(f_n|D,I) \propto \exp\left[\frac{C(f_n)}{\sigma^2}\right], 
\label{hav}
\end{equation}

where $C(f_n)$ is the squared magnitude of the FFT, showing that proper approach to convert  
$C(f_n)$ into probabilities involves first dividing by the noise variance and then exponentiating, 
which suppresses spurious ripples in the power spectrum.

We used a procedure which does not require binning and 
takes into consideration the rotational modulation and decaying 
tail of the  flare \citep{brett88,GL1992,GL1996,2003prth.book.....J,2009MNRAS.tmp.1752V}.

For the analysis of the data for the search of QPOs during the giant flare of SGR 1806$-$20, we applied 
a Bayesian method developed by \cite{GL1992} (hereafter referred to as the GL method) 
for the search of pulsed emission from pulsars in X-ray data, consisting of the arrival 
times of events, when we have no specific prior information about the shape of the signal.
As a particular case of QPO can be considered a periodic signal with some length 
of coherence ($Q\equiv\nu_0/2\sigma, \nu_0 $ is the centroid of the frequency and 
$\sigma $ half width at half maximum),  
i.e. a periodic signal with additional parameters of the oscillation with 
start and end times \footnote{More strictly: coherent transient signal \citep{klis_priv}}.

The GL method for timing analysis first tests if a constant, variable
or periodic signal is present in a data set.

In the GL method, periodic models are represented by a signal folded into trial frequency 
with a light curve shape as a stepwise function with $m$ phase bins per period plus a 
noise contribution. With such a model we are able to approximate a phase folded light curve of any shape. 
Hypotheses for detecting periodic signals represents a class of stepwise periodic models 
with the following parameters: trial period, phase, noise parameter and number of bins ($m$). 
The most probable model parameters are then estimated by marginalization of the posterior 
probability over the prior specified range of each parameter. In Bayesian statistics 
posterior probability contains a term that penalizes complex models (unless there is no significant
evidence to support that hypothesis), hence we 
calculate the posterior probability by marginalizing over a range of models, 
corresponding to a prior range of number of phase bins, $m$, from 2 to 12.
Moreover, the GL method is well suited for variability detection, i.e. 
to characterize an arbitrary shape light curve with 
piecewise constant function $Z(t)$ \citep{2006ASPC..351...73R}.

For the search and detection of QPOs we used a slightly different version of the GL method:
First we determine $Z(t)$ - fitting with a piecewise constant model 
\citep[solid line in the inset panel of the Fig.~\ref{lc},][]{Hutter:07pcregx}

 to characterize
the complex light curve shape in the data set, giant flare decaying tail and 
rotational modulated light curve. Then we subsequently
compare competing hypothesis, i.e. whether the data support a purely piecewise constant
or piecewise constant+periodic model. If there is an indication of the
presence of a periodic signal 
(e.g. odds ratio of competing models exceeding 1, see also \cite{GL1992,GL1996})
 
we determine also the time intervals where 
it has its maximum strength (e.g. amplitude or pulsed fraction) 
via a Markov Chain Monte Carlo (MCMC) approach using 
QPO start and end times as free parameters.

Finally, in the latter case, we estimate parameters
(frequency, phase, amplitude, coherence length of QPO, etc.) 
of the periodic signal with high precision. 
For example, in order to estimate the frequency 
of a periodic signal the posterior probabilities density function used:

\begin{equation}
p(\omega | D, M_m)=\frac{C}{\omega}\int_{0}^{2\pi} d\phi \frac{1}{W_m(\omega,\phi)}, 
\end{equation}
where $C=\left[\int_{\omega_{lo}}^{\omega_{hi}}\frac{d\omega}{\omega}\int_{0}^{2\pi} d\phi \frac{1}{W_m(\omega,\phi)}\right]^{-1}$ and 
$W_m(\omega,\phi)=\frac{N!}{n_1!n_2!\cdot\cdot\cdot n_m!}$ is the
number of ways the binned distribution could have arisen "by chance",
$n_j(\omega,\phi)$ is the number of events falling into the $j\,\rm{th}$
of $m$ phase bins given the frequency, $\omega$, and phase, $\phi$, and $N$ is the total number of photons
\citep[for details, see][]{GL1992}.

\begin{figure}
   \centering
   \includegraphics[width=8.6cm]{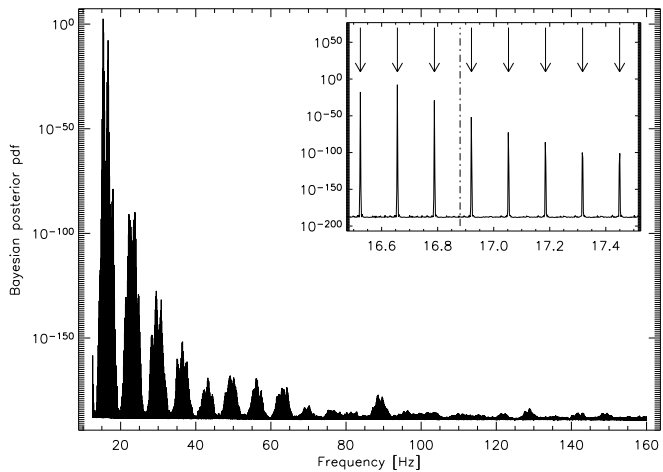}
      \caption{Application of the GL method for blind periodicity search to the complete data set obtained 
by \emph{RXTE PCA} of giant flare 
of SGR\, 1806$-$20 2004 Dec 27 revealed a strong coherent signal at the frequency 
of 0.13219244\, Hz and higher harmonics up to 100Hz.
Bayesian posterior probability density vs frequencies in the range of 12.0-160 Hz is shown. 
The inset panel shows a zoomed part of it around 16.88 Hz. 
Arrows are indicating higher harmonics (from 125 to 132) of the fundamental frequency and
the dash-dotted vertical line shows one of the detected QPO frequency in 
a short time interval (see text, for details).
              }
         \label{FigHarms}
   \end{figure}

First, we applied the GL method as implemented by  \cite{GL1992} to the 2004 Dec 27 giant 
flare whole data set of SGR 1806$-$20 observed by \emph{RXTE PCA}
and started the timing analysis by performing a blind periodicity 
search in the range of 12.0-160 Hz. Naturally, we found a very strong coherent
signal at $\sim$ 7.56s, the pulsation period of the NS, followed by higher harmonics 
up to the 100Hz (see Fig.~\ref{FigHarms}).

   \begin{figure}
   \centering
   \includegraphics[width=7.6cm]{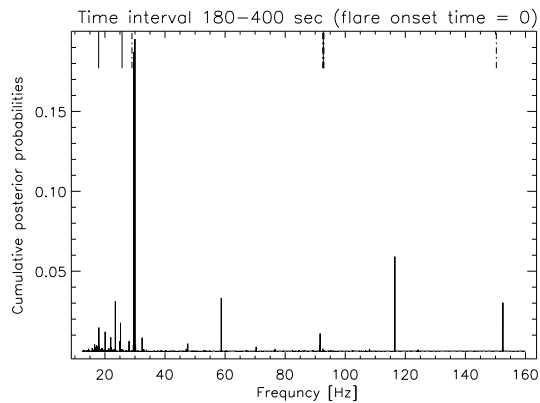}
      \caption{The Bayesian posterior probability density vs. trial frequency of the periodic 
signal for the giant flare data set of SGR 1806$-$20. QPO frequencies already detected by 
averaged power spectrum analysis and also with other mission \emph{RHESSI}
\citep{I2005,WS2006a} are marked as small lines at the top axis. 
Those QPO frequencies are also detected by us. In addition, we have detected several more QPO frequencies 
(21, 59 and 116 Hz) with the Bayesian method, which were also  predicted by  \cite{CBK2009}. 
              }
         \label{pdgm}
   \end{figure}

Next, we divided the data set into 51 rotational cycles (see Fig.\ref{lc}). 
Each of these subsamples were treated as independent data sets. 
We determined the Bayesian probability densities versus trial frequency.  
Final probabilities were derived in two ways; First, as 
the multiplication (the likelihood) of those 
independent Bayesian posterior probability density functions;
and the second one simply summing those independent 
probabilities\footnote{Addition rule of probabilities: 
probability that the QPOs at the given frequency are present 
$\Delta t_1$ or $\Delta t_2$ or both time intervals, while likelihood 
(i.e. summed power spectrum) defines the probability of presence QPOs 
during $\Delta t_1$ and $\Delta t_2$ (i.e. multiplication 
rule of probabilities, see also, Eq.~\ref{hav})} (see Fig.\ref{pdgm}).

This approach also revealed a number of rotational cycles within which the 
probability of the model of periodic signal is significantly higher than the constant one. 
Namely, during time intervals of 183.8$-$191.2, 244.3$-$251.9 and 259.4$-$267.1 seconds, 
from the flare onset, odds ratios of periodic vs piece-wise constant 
models are $\sim$~10, $\sim$~30 and $\sim$~197, correspondingly.  

\begin{figure}
 	\centering
\includegraphics[width=8cm]{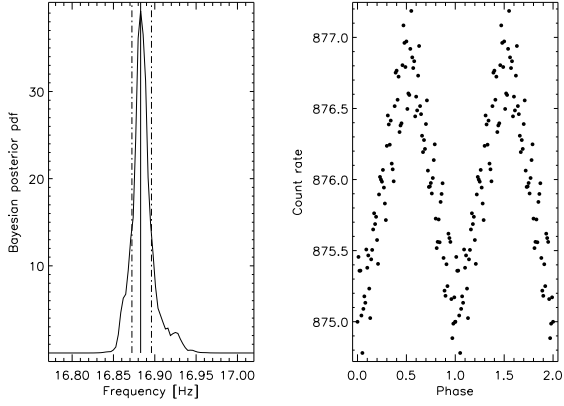}
     \caption{Bayesian posterior probability density vs frequencies in the time 
interval of 259-267 sec from the giant flare onset. By dashed vertical lines are 
indicated the 68\% region of the highest probability density.
 Odds-ratio of periodic vs constant model is $\sim$~200. The right panel depicts 
the phase folded light curve with frequency $f_{QPO}=16.88\rm{Hz}$,it has almost perfect sinusoidal shape 
(for details, see text).
              }
        \label{f34}
  \end{figure}

In order to detect  start and end times oscillations we included also observational data of 
neighboring rotational cycles (where an oscillation  with that frequency was not detected) and 
performed the periodicity search for an expanded time interval with 
additional two free parameters with the MCMC approach with 
Metropolis-Hastings algorithm. As initial values of these 
($t_{QPO start}$ and  $t_{QPO end}$) parameters served start and end times 
of an observation, satisfying the condition: 
$t_{Obs. start} \le t_{QPO start} < t_{QPO end} \le t_{Obs. end}$. 
Starting with the abovementioned initial values can be considered as a 
good strategy, since the detected signal has a higher significance in a subinterval 
of the considered time interval and the fast convergence of the MCMC procedure already provided. 
 This analysis via MCMC revealed even shorter time intervals 
within which periodic signal is stronger. 
The estimates of oscillations frequencies and the corresponding 68\% interval of 
the highest probability densities are presented in the 
Table~\ref{ax} (see, also Fig.~\ref{f34},\ref{fsig}). 

\begin{figure}
 	\centering
\includegraphics[width=8cm]{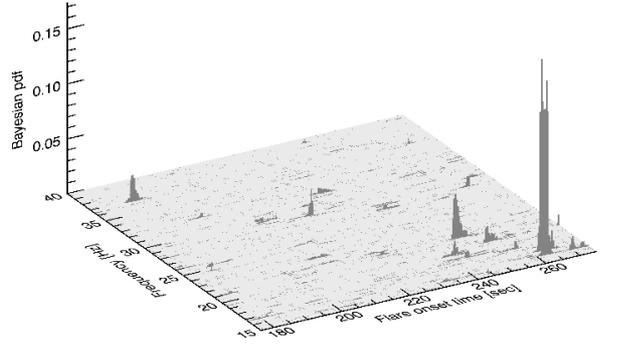}
     \caption{Bayesian posterior probability density vs frequencies and the time of  
the giant flare onset. Clearly seen significant detections of oscillations at 16.9,21.4 and 36.4Hz. 
Large $Q\equiv\nu_0/2\sigma$ values (see, Table~\ref{ax}) indicating on the 
coherent nature of transient oscillations (for details, see text).
              }
        \label{fdyn}
  \end{figure}

\begin{figure}
 	\centering
\includegraphics[width=5.0cm,angle=-90]{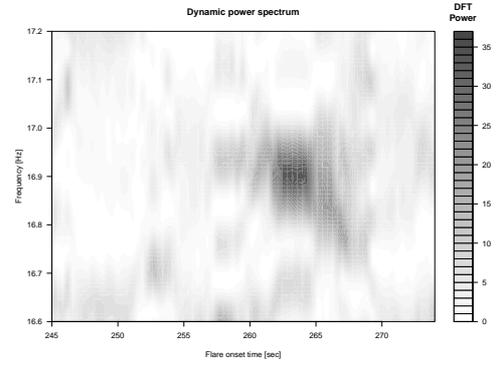}
     \caption{Dynamic power spectrum of giant flare of SGR1806-20 in the frequency 
range of 16.5-17.5 Hz and in the time interval of 250-275sec (see, Table~\ref{ax}). }
        \label{fdyn1}
  \end{figure}

As shown by our intensive simulations, this approach readily detects time intervals 
where periodic signal has highest and lowest\footnote{A time interval 
where odds ratio of the models of a periodic and piecewise constant models 
just exceeds a theoretical minimum, $\mathrm{Odds~ratio~=~1}$, i.e. both models have equal probability to be true.} 
strength. Nevertheless, we performed also time-frequency analysis \citep[see Fig.~\ref{fdyn} 
and \ref{fdyn1}, as well as dynamic power spectrum,][]{1989tns..conf...27V} and
candidate QPOs fitted to a Lorentzian function. We estimated frequency centroids and full widths 
at half maximums. For significant detections they are presented in Table~\ref{ax}. The coherence 
values $Q\equiv\nu_0/2\sigma\sim$60$-$200 are indicating on the transient nature of almost perfect periodic signals.

For those significant detections we have also fitted folded light curves with a template function,
$\psi(t)=\sum_{i=1}^{n} ~A_{i} sin[i \omega_0(t-\psi_{i})]$, truncating the series at the highest harmonic that {\it was}
 statistically significant after performing an F--test \citep[see][for details]{2003ApJ...599..485D}. 
It turned out that oscillations at 16.9Hz and 36.8 Hz adding the second harmonic does not improve 
the fit of phase folded light curve (signifcance levels, accordinglly are 0.71 and 0.16),
while in the case of 21.3Hz it significantly improves it (significance level of 0.0006. 
Thus, oscillations at the frequencies of 16.9Hz and 36.8Hz can 
be considered of sinusoidal shape, and at 21.3Hz not. We obtained similar 
results by analysing also power spectra, i.e. 
fitting spectral profiles of them with {\it sinc} function.

   \begin{table}
      \caption[]{Detected QPO frequencies not reported in the literature \citep{I2005,WS2006a,SW2006}.}
         \label{ax}
	\centering
         \begin{tabular}{cccc}
            \hline\hline
            $f_{QPO}~[\mathrm{Hz}]$ & $\mathrm{Time~intervals}$ & $\mathrm{Lorentzian}$ & $FWHM [\mathrm{Hz}]$ \\
            $\mathrm{(~68\%~credible~region)}$ &  $\mathrm{of~QPOs}^{\mathrm{a}}$ & $\mathrm{centroids (Hz)}$ &  \\
            \hline
%         \begin{tabular}{lccc}
            $16.88^{\mathrm{b}}\ \ ~(16.87-16.90)$ &  259.4$-$267.1 &  16.90$\pm$0.004  &  0.12  \\
            $21.36^{\mathrm{b}}\ \ ~(21.35-21.38) $ &  244.3$-$251.9 &  21.34$\pm$0.003  &  0.35  \\
            $36.84^{\mathrm{b}}\ \ ~(36.83-36.88)$ &  183.8$-$191.2 &  36.88$\pm$0.004  &  0.24  \\
            $59.04\ \ ~(58.58-59.28)$ &  146.0$-$176.2 &  $-$  &   $-$  \\
            $61.26\ \ ~(61.25-61.27)$ &  251.9$-$395.6 &   $-$  &   $-$  \\
            $116.27\ \ \ ~(116.24-116.28)$ &  168.7$-$198.9  &   $-$  &   $-$ \\
%         \end{tabular}
            \hline
         \end{tabular}
\begin{list}{}{}
\item[$^{\mathrm{a}}$] Giant Flare onset time is set to 0
\item[$^{\mathrm{b}}$] Highly significant detection (for details, see text)
\end{list}
   \end{table}

\smallskip

\section{Discussion}

\begin{figure}
   \centering
   \includegraphics[width=7.0cm,clip=]{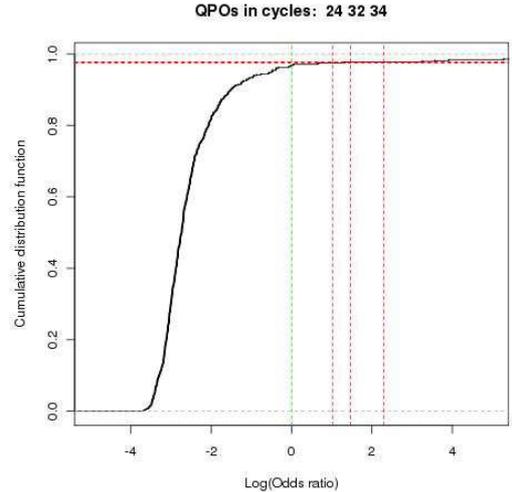}
      \caption{Empirical cumulative distribution function as a function of Odds ratio, i.e. 
the probability of periodic model vs constant one, 
on the base of simulations of QPO frequencies with different amplitudes and noise, 
as much as possible to the observed values in terms of number of registered photons, 
spanned times, etc.. Detected QPO frequencies indicated 
by vertical dashed lines, showing a high significance of detections.
              }
         \label{fsig}
   \end{figure}

We report the detection of oscillations of transient nature from SGR\, 1806$-$20 
giant flare decaying tail recorded by the {\emph RXTE PCA} 
by applying the GL method for periodicity search. We have confirmed the detections of QPOs at frequencies 
(in the range of $12-160 \mathrm{Hz}$) reported
by \cite{WS2006a} and, in addition, we found some more QPOs at  $f_{QPOs}=16.9, 21.4,36.8 \rm{Hz}$ 
with corresponding odds ratios $\sim$~197, $\sim$~30 and $\sim$~10, 
in  shorter time scales, i.e. within individual rotational cycles (see, Fig.\ref{fsig}).

\smallskip

These odds ratios, describing the significance of 
presence of a periodic signal, are sensitive to the frequency search range.
In contrast, detected frequencies of QPOs,  found by locating 
maximums in the posterior probability density function, are insensitive to the prior 
search range of frequencies. We computed the uncertainty of $f_{QPOs}$ at 68\% confidence level 
by using this posterior probability density function. In addition to that, 
we have also estimated the significance of our QPO detection by an empirically 
determined cumulative distribution (see, Fig.\ref{fsig}).

The great variety of neutron star oscillation modes \citep{1985ApJ...297L..37M}
is associated with their global structure as well as their internal
characteristics. Neutron star seismology has already been proposed as a tool
to understand their internal structure \citep{2001MNRAS.320..307K}.
The various types of oscillations carry information about the equation of state, the
thickness of the crust, the mass, radius and even the rotation rate \citep{2010arXiv1005.5228G}.
Still, modeling a truly realistic oscillating neutron star is difficult,
however the potential reward is considerable. This has been demonstrated
from the recent theoretical results for the QPOs which have been
interpreted as magnetoelastic oscillations. These calculations
have shown how the observations constrain both the mass, the
radius, the thickness of the crust and the strength of magnetic
field of these stars \citep{SKS2007,SA2007,CBK2009,DSF2009}.
While one could even set severe constrains in the geometry of the interior 
magnetic field \citep[see][]{2008MNRAS.385.2161S}.

In particular, by studying Alfv\'en oscillations, 
in the stellar interior, in the absence of crust, \cite{CBK2009} 
could reveal the existence of Alfv\'en continua \citep[as suggested by][]{2007MNRAS.377..159L}.
The edges  of this continuum have been used to explain the observed QPOs. It is worth noticing, 
that QPO frequencies found in this work have been  predicted by \cite{CBK2009}.
More recent calculations by \cite{CK2010Pre}
taking into account the presence of crust, i.e. the coupled crust-core oscillations,  
verified the presence of the continuum but in addition they revealed the presence of discrete Alfv\'en 
oscillations as well as some extra discrete modes which can be associated with the oscillations of 
the crust. 
These new results explain the previous published QPO frequencies but also the ones presented here and 
set quite severe constraints in the parameters of the neutron star.

\section{Conclusions}
   \begin{enumerate}
      \item We found new {\bf oscillationsof transient nature} applying Bayesian timing analysis method of the 
decaying tail of the giant flare of SGR\, 1806$-$20 2004 Dec 27, observed by \emph{RXTE PCA}, 
not yet reported in the literature.
      \item   Some of those {\bf oscillations} frequencies ($\mathrm{f}_{QPOs} \sim 17,22,37,56,112 \mathrm{Hz}$) 
are predicted by the theoretical study of torsional Alfv\'en oscillations of magnetars 
\citep[see, Table 2, by][]{CBK2009}, suggesting  $APR_{14}$ \citep{1998PhRvC..58.1804A} 
EoS\footnote{Neutron star models based on the models for the nucleon-nucleon interaction with 
the inclusion of a parameterized three-body force and relativistic boost corrections, 
estimating maximum mass and stiffness \citep[see,also][]{1999ApJ...525L..45H}.}
 of SGR\, 1806$-$20.
      \item These preliminary results are very promising and we plan to extend our 
high frequency oscillations research (both the theoretical predictions as well 
as the observations) to both activity periods, as well as to the quiescent state of SGRs 
and AXPs, as well as to the isolated neutron stars with comparatively smaller magnetic fields.
   \end{enumerate}

\begin{acknowledgements}
      We acknowledge support by the German
      \emph{Deut\-sche For\-schungs\-ge\-mein\-schaft (DFG)\/} through project
      C7 of SFB/TR~7 ``Gravitationswellenastronomie'' This research has made use
of data obtained from the High Energy Astrophysics Science Archive Research Center (HEASARC)
provided by NASA's Goddard Space Flight Center.
\end{acknowledgements}

\bibliographystyle{aa} % style aa.bst
\bibliography{h} % your references Yourfile.bib
\end{document}